\documentclass[12pt]{iopart}
\usepackage{graphicx}
\begin{document}

\rightline{\rm\small IIT-HEP-02/3}

\title[Short title]{Progress in Muon Cooling Research and Development{$^*$}}
\setcounter{footnote}{0}\footnotetext{\leftline{} * To appear in {\sl Proc.\ 4th Workshop on Neutrino Factories based on Muon Storage Rings (NuFact'02)}, Imperial College,
London, United Kingdom, 1--6 July 2002.}

\author{Daniel M Kaplan{\rm \dag\ \ddag\ \footnote[5]{E-mail: kaplan@fnal.gov}}}

\address{\dag\ for the MuCool Collaboration\\ \vspace{.2in}
\ddag\ Physics Division, Illinois Institute of Technology, 3101 S. Dearborn Street, Chicago, Illinois, 60616 USA
}

\begin{abstract}
The MuCool R\&D program is described. The aim of MuCool is to develop all key pieces of hardware required for ionization cooling of a muon beam. This effort will lead to a more detailed understanding of the construction and operating costs of such hardware, as well as to optimized designs that can be used to build a Neutrino Factory or Muon Collider. This work is being undertaken by a broad collaboration including physicists and engineers from many national laboratories and universities in the U.S. and abroad. The intended schedule of work will lead to ionization cooling being well enough established that a construction decision for a Neutrino Factory could be taken before the end of this decade based on a solid technical foundation. 
\end{abstract}




\markboth{}{}

\section{Introduction}

The MuCool Collaboration is pursuing a research and development program on hardware that can be used for cooling of muon beams~\cite{MuCool}. Our recent work~\cite{Alsharoa} focuses on muon ionization cooling~\cite{cooling} for a possible Neutrino Factory~\cite{Geer}, in which an intense, pure, and collimated neutrino beam is produced via decay of a muon beam circulating in a storage ring. The goal for such a facility is $\sim10^{21}$ neutrinos/year (requiring a similar number of muons/year stored in the ring), to exploit the recently established discovery of 
neutrino oscillations~\cite{nu-osc}.

Muon beams at the required intensity can only be produced into a large phase space, but affordable existing acceleration technologies require a small input beam. This mismatch could be alleviated by developing new, large-aperture, acceleration techniques~\cite{Japan-study}, by ``cooling" the muon beam to reduce its size, or both. Given the 2.2-$\mu$s muon lifetime, only one cooling technique is fast enough:  ionization cooling, in which muons repeatedly traverse an energy-absorbing medium, alternating with accelerating devices, within a strongly focusing magnetic lattice~\cite{cooling,Kaplan-cooling}. 

\section{Principle of ionization cooling}

In an ionization-cooling channel, ionization of the energy-absorbing medium decreases all three muon-momentum components without affecting the beam size. By the definition of normalized transverse beam emittance,\footnote{In this expression, for expositional clarity, the effects of possible correlations among the components have been neglected; more rigorously, the normalized transverse emittance is proportional to the square-root of the 4-dimensional covariance matrix of the coordinates $(x,y,p_x,p_y)$ for all particles in the beam.} 
$\epsilon_n\approx\sqrt{\sigma_x\sigma_y\sigma_{p_x}\sigma_{p_y}}/mc$ (where $\sigma_x$, $\sigma_{p_x}$, etc.\ denote the r.m.s.\ deviations of the beam in position and momentum coordinates, $m$ is the particle mass,  and $c$ the speed of light), this constitutes cooling, {\em i.e.}, reduction of normalized emittance. This is so since the reduction of each particle's momentum results in a reduced transverse-momentum spread of the beam as a whole.

At the same time, multiple Coulomb scattering of the muons increases the beam divergence, heating the beam. The equilibrium between these heating and cooling effects is expressed in the following approximate equation for the dependence of $\epsilon_n$ on the distance $s$ traveled through an absorber~\cite{Neuffer2,Fernow}:
\begin{equation}
\frac{{\rm d}\epsilon_n}{{\rm d}s}\ \approx\
-\frac{1}{\beta^2} \left\langle\frac{{\rm d}E_{\mu}}{{\rm d}s}\right\rangle\frac{\epsilon_n}{E_{\mu}}\ +
\ \frac{1}{\beta^3} \frac{\beta_\perp (0.014)^2}{2E_{\mu}m_{\mu}L_R}\,.
\label{eq:cool}
 \end{equation} 
Here, angle brackets denote mean value, $\beta$ is the muon velocity in units of $c$, muon energy $E_\mu$ is in GeV,  $\beta_\perp$ is the lattice beta function evaluated at the location of the absorber, $m_\mu$ is the muon mass in GeV/$c^2$, and $L_R$ is the radiation length of the absorber medium.  (Eq.~\ref{eq:cool} is derived for the cylindrically-symmetric case  of solenoidal focusing, where $\beta_x=\beta_y\equiv\beta_\perp$, by differentiating the expression for $\epsilon_n$ given above.) The first term in Eq.~\ref{eq:cool} is the cooling term and the second is the heating term. To minimize the heating term, which is proportional to the beta function and inversely proportional to radiation length, it has been proposed~\cite{Status-Report} to use liquid hydrogen (LH$_2$) 
as the energy-absorbing medium, giving $\langle {\rm d}E_{\mu}/{\rm d}s\rangle\approx 30\,$MeV/m and $L_R=8.7\,$m~\cite{PDG}, with superconducting-solenoid focusing to give small $\beta_\perp\sim10\,$cm. (Other possible absorber materials are discussed below.)

\begin{figure}
\vspace{-.1in}
\centerline{\scalebox{0.9}{\includegraphics{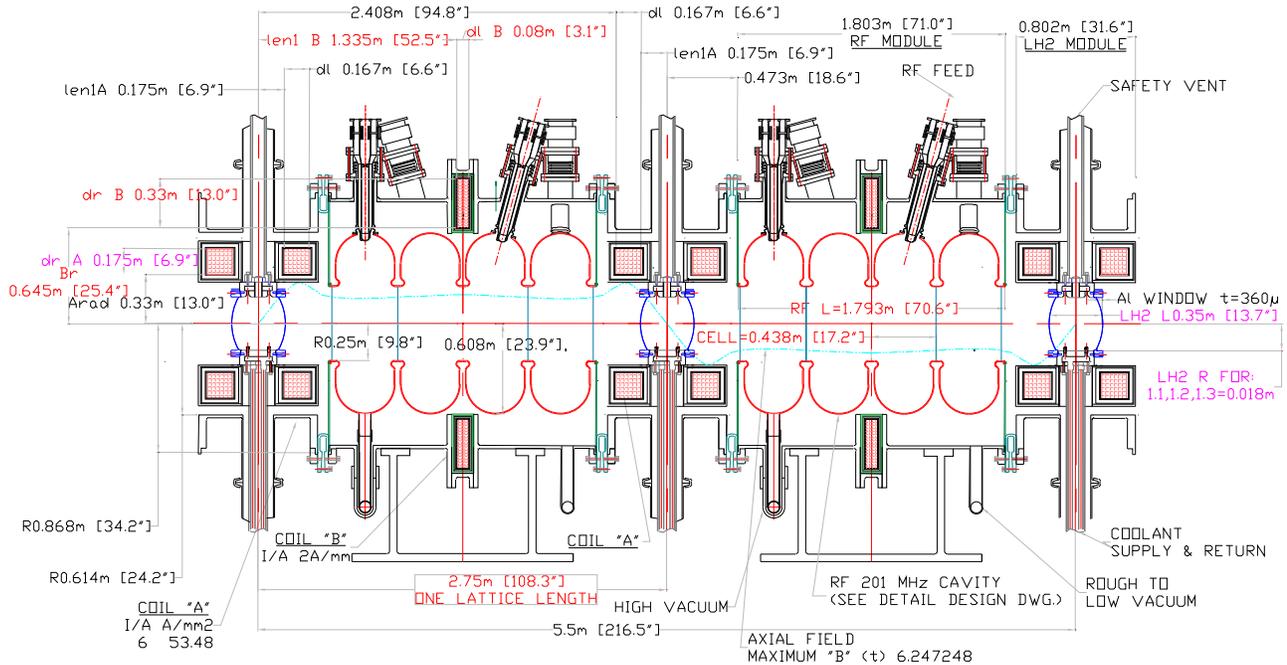}}}
\vspace{-.4in}
\caption{Engineering drawing of a section of an ``SFOFO" ionization-cooling lattice (from U.S. Neutrino Factory Feasibility Study II~\protect\cite{FS2}). Shown in cross section are three liquid-hydrogen absorbers, each enclosed within a pair of ``focusing" solenoids, interspersed with two 4-cavity 201-MHz rf assemblies, each encircled by a ``coupling" solenoid.}
\label{fig:SFOFO}
\end{figure}

Between absorbers, high-gradient acceleration of the muons must be provided to replace the lost longitudinal momentum, so that the ionization-cooling process can be repeated many times. Ideally, the acceleration should exceed the minimum required for momentum replacement, allowing ``off-crest" operation. This gives continual rebunching, so that even a beam with large momentum spread remains captured in the rf bucket. Even though it is the absorbers that actually cool the beam, for typical accelerating gradients ($\sim$10\,MeV/m), the rf cavities dominate the length of the cooling channel (see {\em e.g.}\ Fig.~\ref{fig:SFOFO}). The achievable rf gradient thus determines how much cooling is practical before an appreciable fraction of the muons have decayed or drifted out of the bucket. 

We see from Eq.~\ref{eq:cool}  that the percentage decrease in normalized emittance is proportional to the percentage energy loss, hence cooling in one transverse dimension by a factor $1/e$ requires $\sim$100\% momentum loss and replacement. Low beam momentum is thus favored, because it requires less accelerating voltage and because of the increase of d$E$/d$s$ for momenta below the ionization minimum~\cite{PDG}.  Most Neutrino Factory and Muon Collider beam-cooling designs and simulations to date have therefore used momenta in the range $150-400$\,MeV/$c$. (This is also the momentum range in which the pion-production cross section off of thick targets tends to peak and is thus optimal for muon production as well as  cooling.) The cooling channel of Fig.~\ref{fig:SFOFO} is optimized for a mean muon momentum of 200\,MeV$/c$. 

As a muon beam passes through a transverse ionization-cooling lattice, its longitudinal emittance tends to grow, due to such effects as energy-loss straggling. The six-dimensional emittance (approximately the square of the transverse emittance times the longitudinal emittance) typically is reduced despite this longitudinal heating. However, if not controlled, the longitudinal heating leads to beam losses and thus limits the degree of transverse cooling that is practical to achieve. Cooling lattices with longitudinal--transverse emittance exchange (which can cool  in all six dimensions simultaneously) have been receiving increasing attention and are discussed in detail elsewhere in these Proceedings~\cite{Palmer-ring}. They have the potential to increase Neutrino Factory performance and decrease cost, and are essential to achieving sufficient cooling for a Muon Collider.

\section{Muon-cooling technology development}

An effective ionization-cooling channel must include low-$Z$ absorbers with (if an intense muon beam is to be cooled) high power-handling capability. To achieve low beta at the absorbers requires either high solenoidal magnetic field or high field gradient~\cite{quad}. To pack as much cooling as possible into the shortest distance requires the highest practical accelerating gradient. The MuCool Collaboration has embarked on R\&D on all three of these technologies.

\subsection{High-gradient normal-conducting rf cavities}

An ionization-cooling channel requires insertion of high-gradient rf cavities into a lattice employing strong solenoidal magnetic fields. This precludes the use of superconducting cavities. The lattice of Fig.~\ref{fig:SFOFO} employs normal-conducting 201-MHz cavities, but R\&D is more readily carried out with smaller, higher-frequency devices. Having already embarked on the development of 805-MHz cavities (aimed at a cooling channel for a Muon Collider~\cite{Status-Report}), we have continued to pursue their development, while working out the details of the 201-MHz design in parallel~\cite{Li}. 

Radio-frequency cavities normally contain a minimum of material in the path of the beam. However, the penetrating character of the muon allows the use of closed-cell (``pillbox") cavities, provided that the cell closures are constructed of thin material of long radiation length.  Eq.~\ref{eq:cool} implies that this material will have little effect on cooling performance as long as its thickness $L$ per cooling cell (at the $\beta_\perp$ of its location in the lattice) has $\beta_\perp L/L_R$ small compared to that of an absorber. Closing the rf cells approximately doubles the on-axis accelerating gradient for a given maximum surface electric field, allowing operation with less rf power and suppressing field emission. Two alternatives have been considered for the design of the cell closures: thin beryllium foils and grids of gas-cooled, thin-walled aluminum tubing. As a fall-back, an open-cell cavity design was also  pursued.

So far we have prototyped and tested a 6-cell open-cell cavity, designed at Fermilab,  and a single-cell closed-cell cavity, designed at LBNL, both at 805\,MHz. The tests are carried out in Fermilab's Laboratory G, where we have installed a high-power 805-MHz klystron transmitter (12-MW peak pulsed power with pulse length of 50\,$\mu$s and repetition rate of 15\,Hz), an x-ray-shielded cave, remote-readout test probes, safety-interlock systems, and a control room and workshop area for setup of experiments. The cave also contains a high-vacuum pumping system and water cooling for the cavity. To allow tests of the cooling-channel rf cavities and absorbers in a high magnetic field or high field gradient, a superconducting 5-T solenoid with a room-temperature bore of 44\,cm was constructed by LBNL and installed in Lab G, with two separate coils that can be run in ``solenoid" mode (currents flowing in the same direction) or ``gradient" mode (currents in opposite directions). 

\begin{figure}
\begin{center}
\scalebox{0.5}{\includegraphics*[bb=0 50 600 550,clip]
{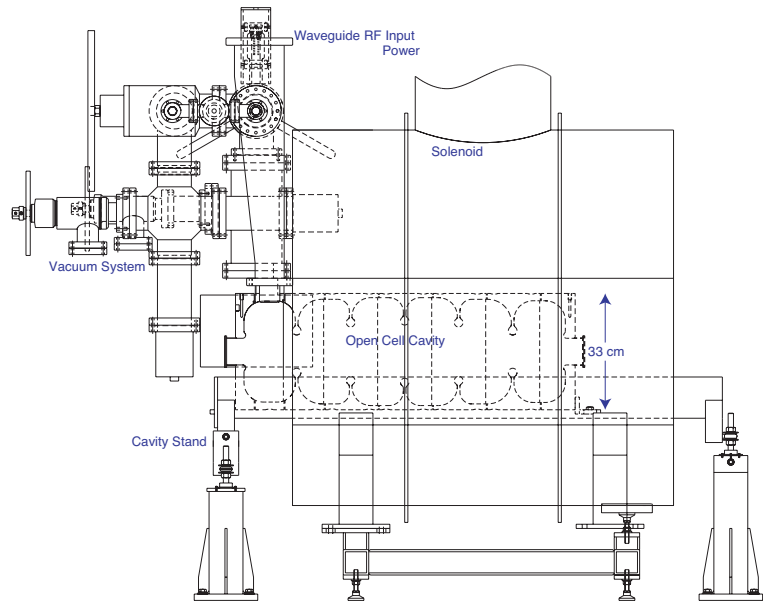}}
\end{center}
\caption{\label{fig:open-cell}6-cell open-cell 805-MHz cavity inserted in Lab G solenoid.}
\end{figure}

The open-cell cavity (Fig.~\ref{fig:open-cell}) was conditioned up to a surface electric field of 54\,MV/m (on-axis accelerating gradient up to 25\,MV/m). Electron dark currents and x-ray backgrounds were found to be large and to scale as a high power of the surface field, $\approx E^{10}$~\cite{Norem}. With a 2.5-T solenoidal field applied, at 54-MV/m surface field, axially focused dark currents ultimately burned a hole in the cavity's titanium vacuum window. This level of background emission would preclude cavity operation in the required solenoidal field. However, for the same accelerating gradient, the pillbox cavity should operate at approximately half the surface field, corresponding to lower background emission by a factor of order $10^{3}$. Furthermore, an analysis of the observed emission rate in terms of the Fowler-Nordheim formalism~\cite{Fowler-Nordheim} implies an enhancement of the emission probability by a factor of order $10^{3}$ compared to that of a smooth, clean surface~\cite{Norem}. This suggests that an R\&D program focused on improving the surface preparation and treatment might reap large improvements.

Tests of the closed-cell prototype have begun. Initial tests up to the design gradient of $\approx$30\,MV/m were carried out successfully with no applied magnetic field. Upon disassembly, no damage to the windows was observed. The thickness of the cavity's vacuum windows precluded measurement of low-energy backgrounds. As of July 2002, a thin window has been installed and reconditioning of the cavity for high-gradient operation has started. So far, the gradients are low and x rays and dark currents have not been seen. Our planned program includes tests of the pillbox cavity with solenoidal field, followed by tests of a variety of surface coatings and cleaning and polishing techniques to identify an optimal approach to building high-gradient normal-conducting cavities for operation in high magnetic fields. In parallel, design studies and prototype tests of beryllium foils and aluminum-tube grids will continue.

\subsection{High-power liquid-hydrogen absorbers}

The development of high-power liquid-hydrogen (LH$_2$) absorbers with thin windows has been a key goal of the MuCool R\&D program~\cite{Kaplan-NuFACT01,Kaplan-windows}. Simulations as well as theory show that scattering in absorber windows degrades muon-cooling performance. To keep this effect to a minimum, the Neutrino Factory Feasibility Study II design~\cite{FS2} calls for aluminum absorber windows of thicknesses given in Table~\ref{tab:lh2}.

\Table{\label{tab:lh2}Specifications of SFOFO-cooling-channel energy absorbers (from U.S. Neutrino Factory Feasibility Study II~\protect\cite{FS2}).}
\begin{tabular}{lccccc}
\hline\hline
 & Length & Radius & Number & Power & Al window  \\
\raisebox{1.5ex}[0pt]{Absorber} &  (cm) & (cm) & needed & (kW) & thickness ($\mu$m)  \\
\hline
SFOFO 1	& 35 & 18 & 16 &æ$\approx$0.3 &	360 \\
SFOFO 2	& 21 & 11 & 36 & $\approx$0.1 & 220 \\
\hline\hline
\end{tabular}
\endTable

The power dissipated per absorber as given in Table~\ref{tab:lh2} is within the bounds of high-power liquid-hydrogen targets developed for, and operated in, a variety of experiments~\cite{targets}.  However, the highly turbulent fluid dynamics involved in the heat-exchange process necessarily requires R\&D for each new configuration. We have identified two possible approaches: a ``conventional" flow-through design with external heat exchanger, similar to that used for high-power LH$_2$ targets, and a convection-cooled design, with internal heat exchanger built into the absorber vessel. The convection design has desirable mechanical simplicity and minimizes the total hydrogen volume in the cooling channel (a significant safety concern), but is expected to be limited to lower power dissipation than the flow-through design. 

Various scenarios have been discussed involving substantially higher absorber power dissipation: 1)~a Neutrino Factory with a more ambitious Proton Driver (4\,MW proton-beam power on the pion-production target instead of the 1\,MW assumed in Study-II) is a relatively straightforward and cost-effective design upgrade~\cite{Alsharoa}, 2)~the ``bunched-beam phase rotation" scenario of Neuffer~\cite{Neuffer-bunch} captures $\mu^+$ and $\mu^-$ simultaneously, doubling the absorber power dissipation, and 3)~a ring cooler~\cite{Palmer-ring} would entail multiple traversals of each absorber by each muon, potentially increasing absorber power dissipation by an order of magnitude. If all three of these design upgrades are implemented, power dissipations of tens of kilowatts per absorber will result. The large heat capacity of hydrogen means that such levels of instantaneous power dissipation are in principle supportable, but much higher average heat transfer would be needed, possibly requiring higher pressure and thicker windows. More work is needed to assess the muon-cooling performance implications.

The large transverse dimensions of the muon beam require large apertures and correspondingly wide absorbers, while the large energy spread of the beam demands frequent rebunching via rf cavities, favoring thin absorbers. These two requirements lead to the oblate shapes of the SFOFO-cooling-channel absorbers indicated in Table~\ref{tab:lh2} and shown in Fig.~\ref{fig:SFOFO}. Since these shapes are wider than they are long, hemispherical windows (which would be thinnest at a given pressure) are ruled out, and we are led to the ``torispherical" window shape. Aluminum alloy is a suitable window material, combining long radiation length with good machinability, weldability, and thermal properties. For an ASME-standard torispherical window~\cite{ASME},  the required minimum thickness is (essentially) $t = 0.885 P D / ES$, with $P$ the maximum pressure differential, $D$ the window diameter, $E$ the weld efficiency, and $S$ the maximum allowable stress, in this case 25\% of the 289-MPa ultimate strength of the 6061-T6 aluminum alloy~\cite{ASME} (the standard alloy for cryogenic and vacuum applications). Taking into account Fermilab's requirement of safe operation at 25-psi (0.17-MPa) differential pressure~\cite{FNAL-safety},  the minimum torispherical window thickness is 760\,$\mu$m of 6061-T6 for the SFOFO 1 absorber (460\,$\mu$m for SFOFO 2), far exceeding the thicknesses called for by the Study-II simulation. To meet the Study-II specifications, we devised a new approach to the design and fabrication of thin windows~\cite{Kaplan-windows}, in which windows of a custom shape and tapered thickness profile are machined out of a solid disk of material using a numerically-controlled lathe, with an integral flange so that no welds are required and $E=1$. We also devised means to test these nonstandard windows and demonstrate that they meet their specifications and satisfy the applicable safety requirements~\cite{MACC}. 

\begin{figure}

\begin{center}
\scalebox{.66}{\includegraphics*[bb=0 50 450 275,clip]{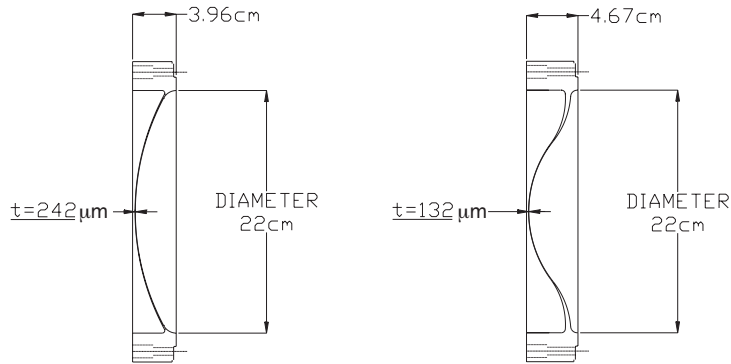}}
\caption{\label{fig:windows}(Left) first and (right) second iteration of custom-shaped and -tapered window design.} 
\end{center}
\end{figure}

Over the past year, as work has continued towards a realistic absorber design, it has become clear that the Fermilab safety code will require external containment of each absorber, to guard against such possibilities as spark damage to a window due to occasional rf-cavity discharges. This doubles the number of windows per absorber, though the containment windows need not be as strong as the absorber windows themselves. We have now developed designs for yet thinner\footnote{That is, thinner at the window center, where the beam is concentrated. Simulation studies have shown that towards the edges, where the new window design is considerably thicker than the old, the muon rate is sufficiently low that the additional window thickness does not degrade cooling performance appreciably.} windows that will allow the Study-II specification to be met even with the additional set of windows per absorber~\cite{Wing}. The old and new window shapes are compared in Fig.~\ref{fig:windows}. We are also exploring the use of new (lithium--aluminum) alloys, such as the 2195 alloy used in the Space Shuttle (80\% stronger than 6061-T6); the resulting thinness of the window may challenge our fabrication techniques, and we will need to certify the new alloy for machinability and high-radiation application. 

\subsection{Other absorber materials}

Other candidate absorber materials include helium, lithium, lithium hydride, methane, and beryllium. All other things being equal, in principle these would all give worse cooling performance than hydrogen. For fixed $\beta_\perp$, a possible figure of merit is $(L_R\,\langle {\rm d}E/{\rm d}s\rangle_{\rm min})^2$ (proportional to the four-dimensional transverse-cooling rate), normalized to that of liquid hydrogen.  Table~\ref{tab:matl} shows that hydrogen is best by a factor $\approx2$, although its advantage could be vitiated if thick windows are necessary. Furthermore, for sufficiently high focusing-current density, lithium lenses could provide substantially lower $\beta_\perp$ than is practical with solenoids, perhaps sufficient to overcome lithium's disadvantageous merit factor. Liquids provide high power-handling capability, since the warmed liquid can be moved to a heat exchanger. On the other hand, the higher densities of solids allow the absorber to be located more precisely at the low-beta point of the lattice. Lithium hydride may be usable with no windows, but means would have to be devised to prevent combustion due to contact with moisture, as well as to avoid melting at high power levels. More work will be required to assess these issues in detail.

\Table{\label{tab:matl}Comparison of ionization-cooling merit factor (see text) for various possible absorber materials~\protect\cite{PDG}.}
\begin{tabular}{lccc}
\hline\hline
 &  $\langle$d$E$/d$s\rangle_{\rm min}$ & $L_R$ & \\
\raisebox{1.5ex}[0pt]{Material} &  (MeV\,g$^{-1}$cm$^{2}$) & (g\,cm$^{-2}$) & \raisebox{1.5ex}[0pt]{Merit} \\
\hline
GH$_2$ & 4.103 & 61.28 & 1.03 \\
LH$_2$	& 4.034 & 61.28 &	1 \\
He	& 1.937 & 94.32 & 0.55 \\
LiH     & 1.94 & 86.9 & 0.47 \\
Li	 & 1.639 & 82.76 & 0.30 \\
CH$_4$	 & 2.417 & 46.22 & 0.20 \\
Be	 & 1.594 & 65.19 & 0.18 \\
\hline\hline
\end{tabular}
\endTable

It has been pointed out~\cite{Kaplan-NuFACT01,MCNote195} that gaseous hydrogen (GH$_2$) at high pressure could provide the energy absorption needed for ionization cooling, with significantly different technical challenges than those of a liquid or solid absorber. Table~\ref{tab:matl} shows that GH$_2$ is actually a slightly better ionization-cooling medium than LH$_2$. In addition, if the hydrogen is allowed to fill the rf cavities, the number of windows in the cooling channel can be substantially reduced, and the length of the channel significantly shortened. Moreover, filling the cavities with a dense gas can suppress breakdown and field emission, via the Paschen effect~\cite{Paschen}. A small business~\cite{MuonsInc} has been formed to pursue this idea, with funding from the U.S. Dept.\ of Energy's
Small Business Technology Transfer program~\cite{STTR}. Phase I, which includes tests of breakdown in gaseous helium at 805\,MHz, 80\,K temperature, and pressures from 1 to 100\,atm,  has been approved. If approved, a follow-on Phase II will explore operation with hydrogen at lower frequency. Successful completion of this program could lead to construction of a prototype gaseous-absorber cooling cell, to be tested at the MuCool Test Area (described next) and perhaps in a future phase of the Muon Ionization Cooling Experiment (MICE)~\cite{MICE}.

\subsection{Test facilities}

To augment the Lab G facility described above, we are building a MuCool Test Area at the end of the Fermilab Linac. This location combines availability of multi-megawatt rf power at  both  805 and 201\,MHz and  400-MeV proton beam at high intensity. Cryogenic
facilities  will be provided for liquid-hydrogen-absorber and superconducting-magnet operation.
The underground enclosure under construction will provide the radiation shielding needed for beam tests of absorber power handling and for high-gradient cavity testing, with the added capability of exploring possible effects on cavity breakdown due to beam irradiation of the cavity walls in a solenoidal magnetic field. 

The MuCool program includes engineering tests of ionization-cooling components and systems, but not an actual experimental demonstration of ionization cooling with a muon beam. Such a cooling demonstration (MICE) has been proposed and is discussed elsewhere in these Proceedings~\cite{MICE}.

\section*{Acknowledgements}

This work was supported in part by
the U.S. Dept.\ of Energy, the National Science Foundation, the Illinois Board of Higher Education, and the Illinois
Dept.\ of Commerce and Community Affairs.


\section*{References}

\begin{thebibliography}{99}

\bibitem{MuCool}
See http://www.fnal.gov/projects/muon\_collider/.

\bibitem{Alsharoa}
M. M. Alsharo'a {\em et al.}, ``Status of Neutrino Factory and Muon Collider Research and Development and Future Plans," FNAL-PUB-02/149-E (July 19, 2002), submitted to Phys.\ Rev.\ ST Accel.\ Beams, arXiv:hep-ex/0207031.

\bibitem{cooling}
A. N. Skrinsky and V. V. Parkhomchuk, Sov. J. Part.\ Nucl.\ {\bf 12}, 223 (1981); \nonum D. Neuffer,
Part.\ Acc.\ {\bf 14}, 75 (1983); \nonum E. A. Perevedentsev and A. N. Skrinsky, in {\sl Proc.\ 12th Int.\ Conf.\ on High Energy Accelerators}, F. T. Cole and R. Donaldson, eds.\ (Fermilab, 1984), p~485; \nonum R. Palmer {\it et al.}, Nucl.\ Phys.\ Proc.\ Suppl.\ {\bf 51A}, 61 (1996).

\bibitem{Geer}
S. Geer, Phys.\ Rev.\ D {\bf 57}, 6989 (1998); \nonum earlier versions of a Neutrino Factory, considered by {\em e.g.} \nonum D. G. Koshkarev, report CERN/ISR-DI/74-62 (1974), \nonum S. Wojicki (unpublished, 1974), \nonum D. Cline and D. Neuffer, AIP Conf.\ Proc.\ {\bf 68}, 846 (1981), and \nonum D. Neuffer, IEEE Trans.\ Nucl.\ Sci.\ {\bf 28}, 2034 (1981), were based on pion injection into a storage ring and had substantially less sensitivity.

\bibitem{nu-osc}
Q. R. Ahmad {\it et al.} (SNO Collaboration), Phys.\ Rev.\ Lett.\ {\bf 89}, 011301 (2002), Phys.\ Rev.\ Lett.\ {\bf 89}, 011302 (2002), and Phys.\ Rev.\ Lett.\ {\bf 87}, 071301 (2001);
\nonum Y. Fukuda {\it et al.} (Super-Kamiokande Collaboration), 
Phys.\ Rev.\ Lett.\ {\bf 81}, 1562 (1998) and Phys.\ Rev.\ Lett.\ {\bf 86}, 5651 (2001);
\nonum B. T. Cleveland {\it et al.} (Homestake Collaboration), Astrophys.\ J. {\bf 496}, 505 (1998);
\nonum R. Davis, D. S. Harmer, and K. C. Hoffman, Phys.\ Rev.\ Lett.\ {\bf 20}, 1205 (1968).
 
\bibitem{Japan-study}
``A Feasibility Study of A Neutrino Factory in Japan," Y. Kuno, ed., available from http://www-prism.kek.jp/nufactj/index.html; \nonum Y. Mori, ``Review of Japanese Neutrino Factory R\&D," \nonum A. Sato, ``Beam dynamics studies of FFAG," \nonum D. Neuffer, ``Recent FFAG studies," \nonum S. Machida, ``Muon Acceleration with FFAGs," and \nonum C. Johnstone, ``FFAG with high frequency RF for rapid acceleration," all presented at this Workshop.

\bibitem{Kaplan-cooling}
An introductory discussion of muon ionization cooling may be found in 
 \nonum D.~M.~Kaplan,
``Introduction to muon cooling,''
in {\sl Proc.\ APS/DPF/DPB Summer Study on the Future of Particle Physics (Snowmass 2001)}, N.~Graf, ed.,
arXiv:physics/0109061 (2002).
\nonum More detailed treatments may be found in  \nonum D.~Neuffer, ``$\mu^+\mu^-$ Colliders," CERN yellow report CERN-99-12 (1999), \nonum  K.~J.~Kim and C.~X.~Wang, Phys.\ Rev.\ Lett.\ {\bf 85}, 760 (2000), and 
\nonum C.~X.~Wang and K.~J.~Kim,
``Linear theory of 6-D ionization cooling,''
in {\sl Proc.\ Snowmass 2001}, {\it op.\  cit.}, SNOWMASS-2001-T502 (2001).

\bibitem{Neuffer2}
D.~Neuffer, in {\bf Advanced Accelerator Concepts}, F.~E.~Mills, ed., AIP 
Conf.\ Proc.\ {\bf 156} (American Institute of Physics, New York, 1987), p~201.

\bibitem{Fernow}
R.~C.~Fernow and J.~C.~Gallardo, Phys.\ Rev.\ E {\bf 52}, 1039 (1995).

\bibitem{Status-Report}
C.~Ankenbrandt {\it et al.}, Phys.\ Rev.\ ST Accel.\ Beams {\bf 2}, 
081001 (1999).

\bibitem{PDG}
K. Hagiwara {\it et al.} (Particle Data Group), Phys.\ Rev.\ D {\bf 66}, 010001 (2002).

\bibitem{FS2}
``Feasibility Study-II of a Muon-Based Neutrino Source," S.
Ozaki, R.~Palmer, M.~Zisman, and J.~Gallardo, eds., BNL-52623, June 2001,
available at  http://www.cap.bnl.gov/mumu/studyii/FS2-report.html.

\bibitem{Palmer-ring}
R. Palmer, ``Ring Coolers: status and prospects" and ``Ring cooler studies," \nonum S. Kahn, ``Simulation of Balbekov ring with realistic fields," \nonum D. Cline, ``Progress in the development of a quadrupole ring cooler
and possible use for neutrino factories and muon colliders," all presented at this Workshop.

\bibitem{quad}
An effort to design quadrupole-focused cooling channels is in progress, but their applicability appears to be limited to the early part of the cooling channel, where relatively large beta functions are appropriate (C. Johnstone, ``Quadrupole channel for muon cooling," presented at this Workshop).

\bibitem{Li}
D. Li, ``201 and 805 MHz cavity developments in MUCOOL," presented at this Workshop.

\bibitem{Norem}
J. Norem {\it et al.}, ``Dark Current Measurements of a Multicell, 805 MHz Cavity," submitted to Phys.\ Rev.\ ST Accel.\ Beams (2002);  \nonum J. Norem, ``RF induced backgrounds at MICE," presented at this Workshop.

\bibitem{Fowler-Nordheim}
R. H. Fowler and L. W. Nordheim, Proc.\ Roy.\ Soc.\
(London) {\bf A119}, 173 (1928).


\bibitem{Kaplan-NuFACT01}
D. M. Kaplan {\it et al.},  ``Progress in Absorber R\&D for Muon Cooling," to appear in {\sl Proc.\ 3rd International Workshop on Neutrino Factory based on Muon Storage Rings (NuFACT'01)}, Tsukuba, Japan, 24--30 May 2001, arXiv:physics/0108027.

\bibitem{Kaplan-windows}
D. M. Kaplan {\it et al.}, ``Progress in Absorber R\&D 2: Windows," in {\sl Proc.\ 2001 Particle Accelerator Conference}, P. Lucas and S. Webber, eds.\ (IEEE, Piscataway, NJ, 2001), p~3888 (arXiv:physics/0108028).

\bibitem{targets}
R. W. Carr {\it et al.}, SLAC-Proposal-E-158, July 1997, and E-158 Liquid Hydrogen Target Milestone Report, http://www.slac.stanford.edu/exp/e158/documents/target.ps.gz (April 21, 1999);  \nonum E. J. Beise {\it et al.}, Nucl.\ Instrum.\ Meth.\ A {\bf 378}, 383 (1996);  \nonum D. J. Margaziotis, in {\sl Proc.\ CEBAF Summer 1992 Workshop}, F. Gross and R. Holt, eds., AIP Conf.\ Proc.\ {\bf 269} (American Institute of Physics, New York, 1993), p~531;  \nonum J. W. Mark, SLAC-PUB-3169 (1984) and references therein.

\bibitem{Neuffer-bunch}
D. Neuffer, ``High frequency buncher and phase rotation," presented at this Workshop.

\bibitem{ASME}
``ASME Boiler and Pressure Vessel Code," ANSI/ASME BPV-VIII-1 (American Society of Mechanical Engineers, New York, 1980), part UG-32.

\bibitem{FNAL-safety}
``Guidelines for the Design, Fabrication, Testing, Installation and Operation of Liquid Hydrogen Targets," Fermilab, Rev.\ May 20, 1997.

\bibitem{MACC}
M. A. Cummings, ``Absorber R\&D in MUCOOL,"  presented at this Workshop.

\bibitem{Wing}
W. Lau, ``Hydrogen Absorber Window Design," presented at this Workshop.

\bibitem{MCNote195} 
R. Johnson and D. M. Kaplan, MuCool Note 195, March 2001 (see
http://www-mucool.fnal.gov/notes/notes.html).

\bibitem{Paschen}
J. M. Meek and J. D. Craggs, {\bf Electrical Breakdown in Gases} (John Wiley \& Sons, 1978),
p~557.

\bibitem{MuonsInc}
Muons, Inc., R. Johnson, Principal Investigator, Batavia, Illinois.

\bibitem{STTR}
See http://sbir.er.doe.gov/SBIR/.

\bibitem{MICE}
R. Edgecock, ``International Muon Ionisation Cooling Experiment: Status and plans," presented at this Workshop; see also http://hep04.phys.iit.edu/cooldemo/.

\end{thebibliography}
\end{document}